# Brain informed transfer learning for categorizing construction hazards

Xiaoshan Zhou[1] | Pin-Chao Liao[1]

[1] Department of Construction Management, Tsinghua University

**Correspondence**
Pin-Chao Liao, No. 30, Shuangqing Rd., HaiDian District, Beijing, 10084, PR China
Email: pinchao@tsinghua.edu.cn

Funding information

National Natural Science Foundation of China, Grant/Award Number: 51878382

**ABSTRACT**

A transfer learning paradigm is proposed for "knowledge" transfer between the human brain and convolutional neural network (CNN) for a construction hazard categorization task. Participants' brain activities are recorded using electroencephalogram (EEG) measurements when viewing the same images (target dataset) as the CNN. The CNN is pretrained on the EEG data and then fine-tuned on the construction scene images. The results reveal that the EEG-pretrained CNN achieves a 9 % higher accuracy compared with a network with same architecture but randomly initialized parameters on a three-class classification task. Brain activity from the left frontal cortex exhibits the highest performance gains, thus indicating high-level cognitive processing during hazard recognition. This work is a step toward improving machine learning algorithms by learning from human-brain signals recorded via a commercially available brain–computer interface. More generalized visual recognition systems can be effectively developed based on this approach of "keep human in the loop."

## 1 INTRODUCTION

Machine learning has experienced a revolution in recent years, driven by neural networks that can complete a wide range of prediction tasks (Rafiei & Adeli, 2016, 2017, 2018; Rafiei et al., 2017). This is particularly true in the vision domain, where machine learning has shown extraordinary performance (Mostafa & Hegazy, 2021; Paneru & Jeelani, 2021). Despite this rapid evolution, machine vision algorithms still perform worse than humans in numerous critical aspects, particularly in tasks involving small-scale training datasets or in highly cluttered realistic scenes (H. Zhou, Xu, et al., 2022). Furthermore, current algorithms can be "fooled" by slightly manipulating images in manners that are undetectable to humans; these lead to misclassifications of objects by machine learning algorithms (van Dyck & Gruber, 2020). Consequently, although machine learning algorithms achieve state-of-the-art performance on some tasks, they do so differently from how humans perform them, and exhibit poor robustness.

As the human brain has evolved to function with optimal efficiency and precision in challenging and ambiguous settings, it provides a natural base of reference for machine learning. Indeed, artificial neural networks outperform other classification algorithms in various vision-related tasks. Moreover, the fields of neuroscience and psychology, and machine learning are getting increasingly intertwined as machine learning algorithms are employed for decoding intentions from recorded brain activities (Houssein et al., 2022). Further, machine learning has taken inspiration from the cognitive functions of the human brain, including memory (Graves et al., 2016), hierarchical processing (Krüger et al., 2013), and replay (van de Ven et al., 2020), to become more intelligent and human-like. Herein, we propose an even more direct method for connecting these two fields, and we investigate whether machine learning algorithms can be used to extract human inner knowledge from recorded human cortical activities when humans and machine learning algorithms perform the same task. Subsequently, we use human knowledge to improve the performance of machine learning algorithms on the task in an explicit manner.

Owing to the complicated workings of the human brain, when allowing machines to learn from human brain activities, a better understanding of the cognition-related functions in the human brain for object recognition is required. Although existing studies have identified saliently activated brain regions based on electroencephalogram (EEG) measurements during hazard identification (Chen et al., 2022; Q. Liu, 2018) (see (Y. Zhang et al., 2019) for a review), the spatial, temporal, and spectral patterns of brain activity for hazard differentiation remain largely unknown. Moreover, with advancements in neuroimaging techniques and the rise in brain–computer interface (BCI) applications, directly decoding intention from brain activities has been explored in the contexts of multiple disciplines. Nevertheless, although several studies have proposed methods for using human knowledge extracted from brain activities to guide machine learning (Fong et al., 2018; Nishida et al., 2020), their data collection and classification methods are highly expensive, thus rendering their methods commercially unavailable. To address these problems, we proposed a method for recording human brain activity using EEG measurements and subsequently employed a parameter-based transfer learning strategy for knowledge transfer between humans and machines. This work builds on previous transfer learning approaches for pretraining neural networks (H. Liu et al., 2021; W. Liu et al., 2018); however, herein, we proposed performing such pretraining using a separate stream of data derived from human brain activity.

Below, we describe our EEG-pretrained transfer learning paradigm in detail, outline an implementation of the technique, and present the results to demonstrate its potential for performing more accurately in an object categorization task. In this study, we trained supervised classification convolutional neural network (CNN) models for three construction hazard categories (namely unclosed electrical compartment, lack of edge protection, and unstable temporary structures). We initialized the model parameters using pretraining on EEG recordings from a human visual cortex viewing the same images. To enable the models to learn the representations in the EEG data, we transformed the brain signals into a two-dimensional (2D) image format using a wavelet transformation. By extracting and incorporating human knowledge of hazard recognition into the CNN models, we hypothesized that our parameter-based transfer learning paradigm would yield better performance on a few-shot classification task. Finally, the models were evaluated for improvements in baseline performance and analyzed to understand which regions of the human brain have greater impacts on performance gains.

## 2 RELATED WORK

### 2.1 Spatial, temporal, and spectral brain activation patterns for hazard recognition

The proposed method works with an underlying hypothesis that cortical activities measured by neuroimaging techniques convey information regarding object categories. Numerous studies have investigated how the human brain responds to objects, such as human faces, body parts, and scenes (Muret et al., 2022; Peelen & Downing, 2017). Additionally, specific functioning regions in the human cortex, such as the temporal lobe for human face processing (Jonas et al., 2016), have been well-characterized for these objects. Nevertheless, for construction hazards, the neuropsychological evidence is limited. Only a few studies have recorded brain activity with EEG signals and used event-related potentials to explore the neurophysiological signatures during the processing of hazards. For example, in previous studies, P300 components were found in the frontal and parietal areas when participants viewed hazardous images (Chen et al., 2022; Q. Liu, 2018; Ma et al., 2014). Although these studies identified critical spatial and temporal characteristics in the brain activity induced by hazardous conditions relative to safe ones, they did not consider that the human brain responds differently to various hazard categories. Nevertheless, previous studies based on field observations and questionnaires have suggested that the recognition of different hazard categories is associated with distinct cognitive processes (Albert et al., 2017; Han et al., 2018). In addition, the findings from the aforementioned EEG studies were acquired via a phase-locked analysis; this approach precludes brain activity appearing with jitter in the latency across trails (Bossi et al., 2020). However, the induced brain activities differing between phases across trials are elicited in trial-based experiments and can only be detected via a time-frequency analysis of the spectral power (Nieuwland et al., 2019).

Several studies have explored the possibility of classifying the brain signals induced by different hazard categories with accuracy above that of chance (Jeon & Cai, 2021, 2022). For example, (Jeon & Cai, 2021) applied multiple machine learning classifiers to predict hazard categories from simultaneously recorded EEG signals, and achieved up to 95 % accuracy in a multiclassification task. These studies have laid the foundation for decoding category information from human brain activities; however, they lack in-depth discussions on the differences in the underlying cognitive disciplines for different hazards. Consequently, artificial intelligence cannot borrow such insights from or take advantage of the inner workings of the human brain.

To gain a deeper understanding of the functional mechanisms of the human brain when recognizing different construction hazards, a previous study recorded participants' neural activities in the prefrontal cortex using functional near-infrared spectroscopy, and applied machine learning algorithms to predict which category of hazards was being recognized from the recorded hemodynamic responses (X. Zhou et al., 2021). Their findings suggested that electricity-related hazards induce a more instinctive judgment, whereas structure-related hazards require episodic memory retrieval from long-term memory. Although this study revealed important insights into the hazard recognition cognitive process, it recorded the brain activity only in the prefrontal area. Evidence from neuropsychology studies suggests that additional posterior regions, such as the parietal cortex, are responsible for actively storing and manipulating visual information (Champod & Petrides, 2007). Studies from smart EEG-based wearables in construction empirically found that posterior regions, such as the parietal and occipital cortices, yield better machine learning performance for monitoring human cognitive states (Noghabaei et al., 2021; D. Wang et al., 2017). These studies suggest that a thorough investigation of the regions of the entire brain is necessary to select the optimal channel(s) for extracting human knowledge of hazard recognition. In addition, the temporal resolution of hemodynamic responses is low; a delay of several seconds degrades accurate attributions for specific cognitive functionality (Bauernfeind et al., 2011).

Therefore, in this study, EEG was chosen as the recording modality owing to its high temporal resolution in capturing fast fluctuations in brain activity on a millisecond time scale (Bressler, 2002). The entire brain area was recorded with a 32-channel electrode cap to investigate which regions of the brain had greater impacts on the performance of the proposed method. Time-frequency analyses were conducted on the collected EEG signals to capture the spectral characteristics of the brain activity. The brain areas, time-variant activation strength, and triangulation with specific frequency bands are expected to provide greater insights into the inner workings of the human brain in recognizing different hazard categories.

## 2.2 Decoding hazard category information from brain activities via brain–computer interfaces

BCIs are the best-known application of neuro-engineering and aim to provide a direct communication pathway between the human brain and external facilities. Most studies have used BCIs to decode movement intentions by classifying brain signals generated by mental imagery tasks (Feng et al., 2021; Hassanpour et al., 2019), thereby enabling direct control of rehabilitation devices for disabled people (Burns et al., 2020; Z. Yang et al., 2018). BCIs are also widely used to diagnose disorders (Bilgen et al., 2020; Nogay & Adeli, 2020; Olamat et al., 2022), recognize emotions (Cai et al., 2022), and assess mental statuses (Jebelli et al., 2017). However, the brain signals in the aforementioned tasks are spontaneous. More recently, BCIs have been extended to analyze changes in the brain activity evoked by external visual stimuli.

A typical series of steps for establishing a BCI includes signal acquisition, preprocessing, feature extraction, classification, and control via an interface (Luis & Gomez-Gil, 2012). Despite the existing BCI paradigms, robust feature extraction and classification are critical steps as they significantly influence the performance of BCIs. EEG studies in construction typically select time-, frequency-, and time–frequency-domain features of the EEG signals as inputs to machine learning classifiers, including the linear discriminant analysis, support vector machine, and k-nearest neighbor (Jeon & Cai, 2021; Saedi et al., 2022). Owing to the high dimensionality of the EEG features, methods of feature dimensionality reduction (such as principal component analysis and t-distributed stochastic neighbor embedding) have begun to be used to learn more discriminative EEG features (Jebelli et al., 2018; Jeon & Cai, 2021). Additionally, automated machine learning approaches have been utilized to tune hyperparameters for optimized classifiers (Jeon & Cai, 2021). Although impressive performance has been achieved by these machine learning classifiers, the methods in these streamlined studies pose high technical barriers. In particular, they require expertise regarding the features of EEG signals and the nature of the investigated cognitive task; this could discourage interdisciplinary research at the intersection of neuroscience and construction safety.

To address the issues of handcrafted EEG features, an end-to-end analysis approach for signal classification enabled by deep learning has emerged recently; it alleviates the need for manual feature extraction and provides exceptional classification performance in various cognitive tasks (Craik et al., 2019; Schirrmeister et al., 2017). The paradigm of transfer learning has been explored in these studies to take advantage of established computer vision models in the literature proven to work well in EEG signal classification (K. Zhang et al., 2020). Although previous studies have explored multiple cognitive tasks, few have considered construction hazards. Thus, further research is required to investigate the possibility of employing transfer learning to classify the brain signals induced by different hazard categories. To transform the raw EEG signals into inputs that can be fed into CNNs, this work takes inspiration from previous studies wherein wavelet transformation was conducted to transform raw brain signals into the time–frequency domain to obtain spatial-spectral-temporal EEG representations (Hajinoroozi et al., 2016; Tabar & Halici, 2017).

## 2.3 Using human brain activity to guide machine learning

This work contributes to the growing body of literature suggesting that additional "side-stream" data sources can be leveraged to improve machine learning algorithms. Although measures of human behavior have been used extensively to guide machine learning via demonstration learning (C.-J. Liang et al., 2020), active learning (Kim et al., 2020), structured domain knowledge (Dash et al., 2022), and discriminative feature selection (Deng et al., 2015), this study proposes that measures of the internal representations employed by the human brain can be directly harnessed to guide machine learning. We argue that this approach opens a new avenue for accelerating research in the intersection of machine learning and neuroscience.

Previous studies have constrained machine learning models via human brain activity. One work introduced a method for weighing the importance of each training sample to machine learning algorithms according to the effort exerted by the human brain to recognize each sample (Fong et al., 2018). They measured human voxel responses to images using functional magnetic resonance imaging (fMRI); their results revealed more consistency in the decision boundary between the machine learning classifiers and human brain representations by the weighted training method.

Furthermore, recent advances in machine learning have focused on improving the learning of discriminative brain activity of manifold visual categories from EEG recordings evoked by visual object stimuli, thereby enabling machines to employ a mapping from images to brain-like features for improved automated visual classification (Spampinato et al., 2017). One work implemented transfer learning to incorporate feature representations learned from human brain activation patterns recorded by fMRI and combined with deep visual features extracted from the images to improve the performance of machine learning in an audiovisual pattern recognition task (Nishida et al., 2020). However, the studies using fMRI modality for brain activity recording suffer from prohibitive costs and poor portability; thus, they can be used only in laboratory experiments. In this study, we demonstrated the potential of a commercially available EEG-based BCI for effectively extracting and incorporating human knowledge from recorded brain activities into machine learning models. The proposed method collectively uses the spatial, temporal, and spectral features of activities, i.e., beyond the features extracted from a single frequency band in previous studies (W. Liu et al., 2018; Nicolae et al., 2020; Nogay & Adeli, 2020), thereby improving the knowledge transfer between humans and machines in an explicit and comprehensive manner.

## 3 METHODOLOGY

### 3.1 Electroencephalogram (EEG) data acquisition

The EEG data used for the machine learning experiments presented in this work are a subset of the data from a published study on construction hazard recognition (Chen et al., 2022). All EEG data were collected using a 32-channel electrode cap (according to the 10–20 international system). Although we recorded EEG activity for 70 participants in this study, we used only the brain activity from one participant. The experiment was approved by the Department of Civil Engineering of Tsinghua University.

### 3.2 Data set

The original dataset for the participant consisted of 120 color images of real construction scenes; the participants viewed the images while their brain activities were recorded (see (Chen et al., 2022) for additional details on the dataset). The visual stimuli involved multiple hazard categories, including overhead power lines, unprotected electrical panels, open electrical compartments, suspended platforms without edge protection, unstable temporary structures (such as scaffolding and frameworks), safety helmets not being used, obstacles on the floor that can cause tripping, and inappropriate storage of chemicals susceptible to fire and explosions. Because electric leakage, lack of edge protection, and structural instability are the most common hazards and precursors of accidents at construction sites (Fang et al., 2020; Han et al., 2018), we used a subset of these three hazard categories, with each category consisting of ten samples (see Fig. 1b for these images). These images were used as both visual stimuli for the EEG data collection and training data for the machine learning experiments.

### 3.3 EEG data preprocessing

The EEG signals were initially subjected to bandpass filtering (0.1–40 Hz). Subsequently, possible artifacts (such as heartbeats, eye movements, and respiration responses) were assessed and removed using an independent component analysis (Makeig et al., 1996). Next, the EEG signals were segmented into a [–200 ms, 1000 ms] epoch on a trial basis, with a [–200 ms, 0 ms] baseline (0 ms denoted the stimulus onset). Furthermore, epochs with values larger than ±100 μv for any recording electrode were rejected to avoid artifact contamination. Offline analyses of the EEG data were conducted using the FieldTrip toolbox (Oostenveld et al., 2011) in MATLAB (version R2019a). A previous study (Chen et al., 2022) provides additional details regarding the EEG data acquisition and preparation not directly related to the machine learning experiments described in this work.

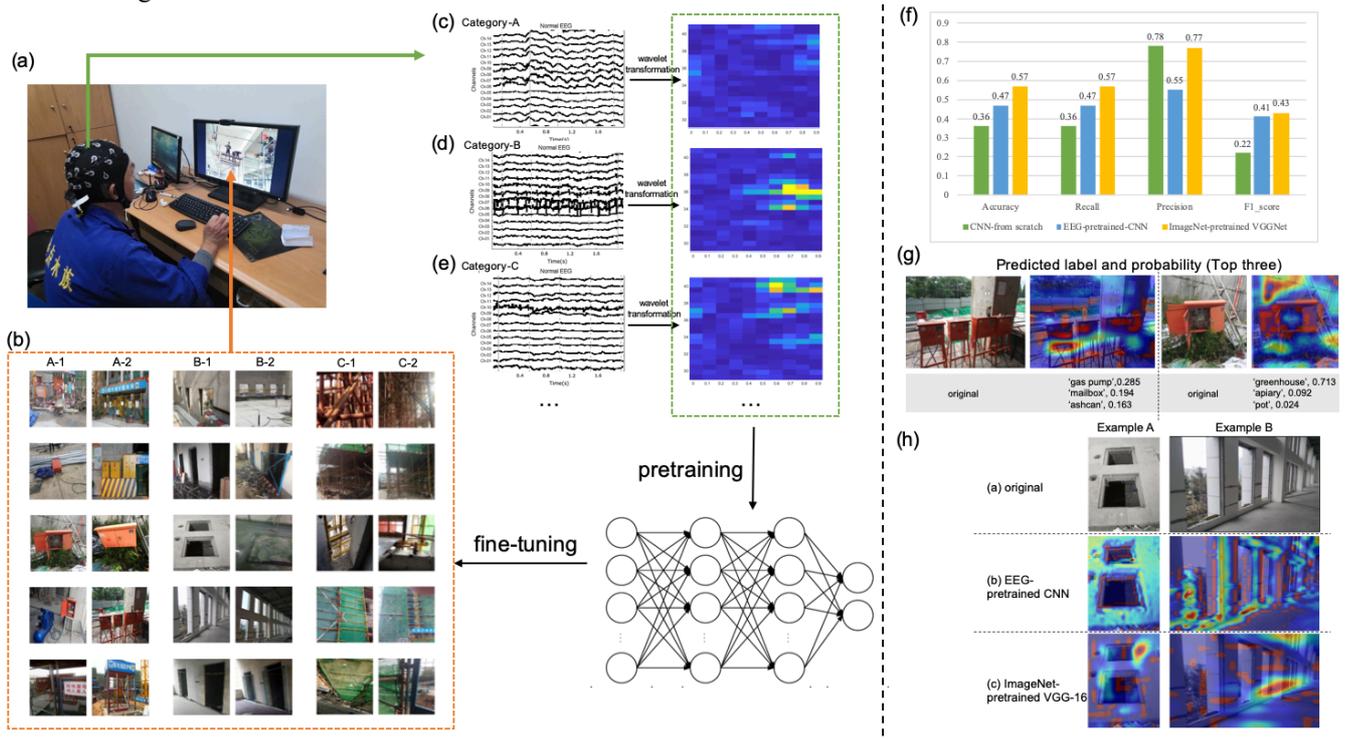

FIGURE 1. (a) Neuropsychological experiment of a construction hazard recognition task with collected electroencephalogram (EEG) data. (b) visual stimuli. Letters show hazard categories (A: electric leakage, B: lack of edge protection, C: structural instability) and numbers show conditions (1: hazardous, 2: safe). (c)–(e) Examples of raw EEG signals and time-frequency maps of EEG representations after wavelet transformation induced by three hazard categories. (f) Classification performance metrics. (g) Gradient-weighted class activation mapping (Grad-CAM) localizations for the "electric leakage" category for ImageNet-pretrained visual geometry group (VGG)-16 and its predicted labels with top three probabilities. (h) Grad-CAM feature heatmaps that highlight important regions for the model prediction and source construction scene image for the "lack of edge protection" category. The original images and their corresponding feature heatmaps are shown from top to bottom. The red part that gathers inward to the blue part indicates that the model pays particular attention to this area.

### 3.4 EEG time-frequency representation calculation

All of the raw EEG signals were transformed into the time–frequency domain (Tallon-Baudry et al., 1996) to serve as 2D inputs (see Figs. 1c–e) subsequently fed into the CNN. Following the previous approach that converted time-series EEG signals into the time–frequency domain to simultaneously preserve the temporal and spectral features of EEG data (Harada et al., 2020), the time–frequency representations of EEG data were generated using the Hanning taper. A 500-ms sliding window with 50-ms time steps was applied. The 0–0.9 s post-stimulus period was chosen for the period of interest, in line with previous findings indicating relationships between brain activities and construction hazard-related visual information processing (X. Zhou, Liao, et al., 2022). The time–frequency analysis of the EEG data was performed using the FieldTrip toolbox in MATLAB.

### 3.5 Experimental design

Following the CNN structure designed for transformed EEG image classification (Xu et al., 2020), we employed a CNN learning framework comprising an input layer, three convolution layers with rectified linear units (ReLus), max-pooling with two fully connected layers, and a softmax output layer for classification (see Table 1 for the details of the CNN architecture).

Keras (version 2.4.3) with Tensorflow (version 2.3.1) was used as the backend for the experiments. When learning on the EEG data, all of the parameters of the CNNs were randomly initialized from Gaussian distributions and trained with a batch size of one image instance. The convolution stride in those layers was performed in each dimension, and the "SAME" mode was used for the spatial padding. After the ReLu activation, the pooling layer filter had a side length of 0, filled with a zero, and a moving step size of 2. Adamax optimization (Kingma & Ba, 2014) with a learning rate of 0.0001 was used, and categorical cross-entropy was used to estimate the loss.

The models were tested using five-fold cross-validation. The full data were split into five equally-sized portions. Out of the five portions, four were used to train the CNN model, whereas the remaining one was used for testing and estimating the classification performance. This procedure was repeated five times to ensure that every portion was used exactly once for testing. Because the dataset was very small, up to 20 epochs were trained and the training could stop early to avoid overfitting, i.e., as soon as the loss on the validation set no longer decreased in three epochs. Subsequently, the CNN was fine-tuned on the construction image dataset. The training-testing splits for the five-fold cross-validation were the same for the EEG images and construction images.

For the baselines, we implemented the same CNN architecture on the construction image dataset from scratch. In addition, we used a mainstream pretrained model, i.e., the visual geometry group (VGG)-16 (Simonyan & Zisserman, 2014). This CNN has a deeper 16-layer structure. These layers contain a series of convolutional layers followed by a single max-pooling layer. The VGG model was acquired as a pretrained model from the ImageNet dataset, which consists of more than 1.2 million 256 × 256 natural images belonging to 1000 classes (Deng et al., 2009). Examples of object categories in ImageNet include "bonnet," "chainlink fence," "castle," and "garbage truck." Because ImageNet contains numerous object categories, it is commonly used as a source dataset for implementing transfer learning.

To train the VGG-16 model, we followed a customary approach from a previous study (Oquab et al., 2014). The top fully connected layer of the VGGNet was replaced with a new fully connected layer with 512 activation units on top, followed by a final layer with a softmax activation function to output the predictions from the three preexisting categories in our dataset. In the fine-tuning process, the early layers were kept fixed. This was because the earlier layers of the CNNs generate more generic features that can be used despite the data distribution. In contrast, higher-level layers of the CNN can devote more representational power to features more specific to differentiating between the task-related object categories. Therefore, we trained only the newly introduced layers with an Adamax optimizer (Kingma & Ba, 2014) with a learning rate of 0.0001.

**TABLE 1.** Details of each layer in the proposed convolutional neural network (CNN) model

| Layer Type | No. of filters | Kernel size | Output shape | No. of trainable parameters |
|---|---|---|---|---|
| Input | / | / | 150 × 150 | / |
| Conv2D | 16 | (3, 3) | 148 × 148 | 448 |
| MaxPooling | 16 | (2, 2) | 74 × 74 | 0 |
| Conv2D | 32 | (3, 3) | 72 × 72 | 4640 |
| MaxPooling | 32 | (2, 2) | 36 × 36 | 0 |
| Conv2D | 64 | (3, 3) | 34 × 34 | 18496 |
| MaxPooling | 64 | (2, 2) | 17 × 17 | 0 |
| Flatten | / | / | 18496 | 0 |
| Dense | / | / | 512 | 9470464 |
| Dense | / | / | 3 | 1539 |

## 3.6 Visualization via gradient-weighted class activation mapping (Grad-CAM)

Inspired by (Melching et al., 2022) and (Y. Yang et al., 2022), gradient-weighted class activation mapping (Grad-CAM) (Rs et al., 2020) was used to interpret the CNN models. Grad-CAM provides visual explanations for deep neural networks via gradient-based localization. It can generate a localization map in which important regions in the image are highlighted, thereby indicating what the network depends on to predict the category. Hence, it provides valuable insights into the interpretability of networks, particularly in diagnosing the failure modes of the networks, as it helps explain seemingly unreasonable predictions by localizing discriminative image regions.

## 4 RESULTS

Because the brain activation powers induced by each hazard category did not fit normal distributions (Lilliefors test, $p < 0.05$), a Kruskal–Wallis test was conducted to investigate the differences in the brain activation patterns induced by the three hazard categories. Numerous studies on EEG decoding have focused on three brain regions: the frontal cortex, posterior parietal cortex, and temporal cortex (Linden et al., 2012). This is partly because these three regions are extensively associated with visual information processing during object recognition. In this study, the statistical results indicated that the left frontal cortex was particularly engaged in the hazard recognition process, with different activation patterns when recognizing different hazards (Fig. 2). This is consistent with the findings of previous neuropsychological studies on construction hazard recognition in that the left frontal cortex is actively involved in enabling a more hazard-related semantic analysis (Jeon & Cai, 2021; X. Zhou et al., 2021). The results obtained herein align well with previous cognitive theories that the left frontal cortex serves as a vital region for object vision in remembering "what" an object is (Wilson et al., 1993). Hence, activities recorded from the F3 electrode were considered as better at capturing the semantic information of the hazard categories encoded in the internal representations of the human brain, and were chosen for the following machine learning experimentation.

As shown in Figs. 1c–e, we found that the spectral power in recognizing structure-related hazards was most pronounced in the gamma band, whereas edge protection-related hazards induced stronger beta-band oscillatory activities. The related cognitive functions are discussed below.

The CNN achieved an average accuracy of 33 % when pretraining the EEG data. We used the accuracy, precision, recall, and F1-score to evaluate the training strategies (Fig. 1f). Evidently, pretraining on the EEG data led to a large improvement in the models relative to random initialization. These results suggest that improvements in the network parameters may be able to compensate for the limited training samples. They also imply that some category information carried by brain activity may already be latently captured in the CNN parameters.

Despite a relatively higher classification accuracy, a further visualization analysis implied that the ImageNet-pretrained CNN might not capture the correct features to depend on when making predictions. For example, it may depend on false cues to make a judgment or be fooled by a highly cluttered environment. As shown in Fig. 1g, VGG-16 mistakenly ascribed the electrical compartment to a gas pump rather than recognizing the dominant object "electrical compartment." It assigned a high probability (>0.7) to "greenhouse" when green plants existed in the environment. For the category of "lack of edge protection," the VGG-16 pretrained on ImageNet predicted the labels of Examples A and B in Fig. 1h with the highest probability as "maze" and "prison," respectively. In contrast, although the key element—railway—was absent in the hazardous conditions, the CNN pretrained on the EEG data could correctly localize the edge, thus indicating that the EEG-pretrained CNN was fairly robust to scene conditions and visual clutter.

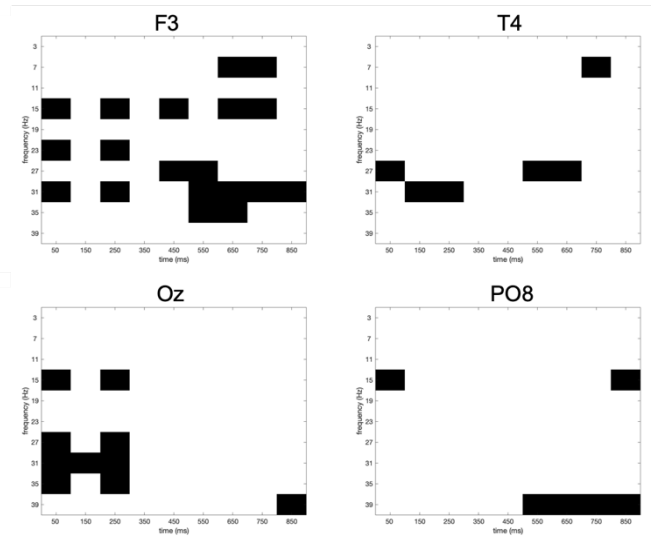

FIGURE 2. Results comparing EEG powers induced by three hazard categories in multiple electrodes for each time-frequency window (time: 0–100, 100–200, 200–300, 300–400, 400–500, 500–600, 600–700, 700–800, and 800–900 ms; frequency: 2–4, 6–8, 10–12, 14–16, 18–20, 22–24, 26–28, 30–32, 34–36, and 38–40 Hz). Blocks filled in black denote corresponding $p < 0.05$.

## 5 DISCUSSION

Our results provide strong evidence that information measured directly from the human brain can help a machine learning algorithm perform better and more reasonable classifications. The implications from decoding object category-related information from EEG signals for inclusion in computer vision methods are tremendous. First, identifying neuropsychological signatures for visual categorization may provide meaningful insights into the functionality of human visual perception systems. For example, the results in this work indicated that recognizing a "lack of edge protection" was associated with stronger beta-band activity in the left frontal cortex. This may mediate a more powerful cognitive control for maintenance of the current cognitive state, i.e., beyond foreseeing a potential falling (Engel & Fries, 2010). In contrast, stronger gamma-band oscillatory activity was elicited when the participant perceived an "unstable structure," thus suggesting the involvement of higher-order sensory processing (R. W. Y. Wang et al., 2020). Although gamma-band activity was initially considered for perceptual processes (Uhlhaas et al., 2009), high-frequency oscillations are also responsible for cognitive functions, such as working memory and attention (Jensen et al., 2007). Recognizing structural instability requires rule-based semantic analysis and episodic memory retrieval (X. Zhou et al., 2021), which may lead to greater engagement with gamma-band oscillatory control for visuospatial working memory maintenance (Jokisch & Jensen, 2007) and selective attention modulation (Shomstein & Yantis, 2004).

Second, the identified neuropsychological evidence will considerably advance the performance and generalizability of BCI-based applications. Currently, the efforts required for EEG experimental setups are massive because conductive glue must be used on the participant's scalp to force the electrode impedance below 20 Ω to collect high-quality EEG signals. Analyzing the high-dimensional data acquired from dozens of electrodes also leads to excessive learning time and overfitting of the training data. Thus, the brain area from which hazard category information can be decoded most efficiently must be identified. The findings from this study provide important references for optimal recording site selections for BCI devices. In contrast to electrodes placed over the entire scalp, this approach considerably reduces the extravagant development costs of a BCI application and makes it commercially available. Additionally, it renders the BCI more user-friendly for conversion into a real-world application.

In addition, the BCI enables a new form of brain-based image labeling. This opens a new avenue to replace, or at least supplement, human manual annotation by decoding hazard-category labels from induced human cortical activities. Although this study focused on construction hazard object categorization, the framework described here does not need to be specific to the construction context. Moreover, effectively pretraining the CNN modes on human brain activity data will dramatically change the way object classifiers are developed. Machine learning can benefit from this form of "keep human in the loop." As the learning from human brain activity improves the performance of the machine learning classifiers, it can also reduce the need for a large-scale dataset for training the neural networks, thereby collectively contributing to the alleviation of manual efforts in image labeling.

Finally, this study also demonstrates the potential of a BCI for enabling intuitive communication of human perceptions for hazard-category information; these can replace verbal communication or button-press interfaces in traditional device design. This communication channel lays the foundation for human-machine collaborations for ubiquitous hazard identification by enabling semi-automated navigation and perception systems to be better guided by human cognition.

The initial implementation of the proposed paradigm has several limitations, which can be addressed through further research. First, the structure of the CNN used in this study was borrowed from a previous study that also conducted EEG image classification. It is a lightweight network compared with complex solutions, and was chosen to reduce computational costs (Arco et al., 2022; Y. Liang et al., 2022). Future studies should use a neural architecture search for an automatic design of an appropriate (typically small) network structure to replace the design constructed by human experts (Xue et al., 2021).

Second, the EEG data used in this study were collected when participants performed a binary discrimination task, that is, the presence or absence of hazards. Researchers could use a multiple-choice design in the experiment to make the participants undertake a more comprehensive scene understanding and risk assessment. We can gain a better understanding of how the human brain functions during hazard recognition by monitoring the regions that are more activated during these complex cognitive functions. Moreover, we can provide feedback in the experiment to help participants learn to avoid making mistakes. This can help prevent the introduction of noise from false trials into the recorded brain signals (Grill-Spector et al., 2000). More generally, using the method developed in this work, we can test whether the recorded brain activity may encode superior domain knowledge and further improve the performance of the machine learning algorithms.

Third, the proposed method was tested on EEG data acquired from one participant, and the performance gain of the EEG-pretrained CNN model was observed. This, combined with the identified optimal recording electrode, provided a huge practical advantage in reducing the efforts required for collecting EEG data. In future, researchers can extend this work to investigate individual differences in recognition processes using the developed method. This can also contribute to the development of personalized BCI-based services and systems to provide a better experience for BCI users.

Finally, with regard to efforts to acquire the EEG dataset, numerous EEG datasets are publicly available, and researchers are encouraged to test our paradigm on these datasets. Researchers from computer science and neuropsychology are also highly encouraged to collaborate to record brain activities in neuropsychological experiments for various object categorization tasks, and to use the brain activation patterns encoded in the recorded data to advance the development of advanced computer vision models. In addition, we will publicly release the EEG dataset used in this study in the future.

## 6  CONCLUSION

We proposed a transfer learning paradigm for regularizing initialization for CNN models on a categorization task by pretraining CNNs on human brain activities recorded when humans perceived the same images as machine learning algorithms. We selected EEG signals for the brain signal recording owing to their high temporal resolution, portability, and commercial viability (Liao et al., 2022). The transfer learning paradigm was utilized as it has been proven effective in multiple cross-modality image settings in previous studies (Jia et al., 2021). To evaluate the performance of the proposed EEG-pretrained method, we compared it with the performance of the randomly initialized and ImageNet-pretrained methods on a three-class construction hazard categorization task with a small-scale dataset. Our results demonstrated that the EEG-pretrained method outperforms the other methods in terms of classification accuracy and interpretability as revealed by visualization. Therefore, pretraining CNN models on EEG data provides an effective method for extracting and incorporating human knowledge of hazard recognition from recorded brain activities into CNN models. Although this approach has some limitations, we hope that this work will serve as a springboard for the increasingly important discussions on transferring human visual capabilities to machine learning algorithms and developing human-like artificial intelligence.

## ACKNOWLEDGMENT

This study was supported by the National Natural Science Foundation of China (No. 51878382).